\documentclass[showkeys,showpacs,10pt]{revtex4-1}
\usepackage{graphicx}
\usepackage{color}
\usepackage[english]{babel}
\usepackage{amsmath,latexsym,amssymb}

\begin{document}
\title{Relativistic heat conduction: the kinetic theory approach and comparison with Marle's model}
\author{A. R. M\'endez and A. L. Garc\'ia-Perciante}
\affiliation{Departamento de Matem\'aticas Aplicadas y Sistemas,\\ Universidad Aut\'onoma Metropolitana- Cuajimalpa,\\ 01120, Cuajimalpa, M\'exico}
\begin{abstract}
In order to close the set of relativistic hydrodynamic equations,
constitutive relations for the dissipative fluxes are required. In
this work we outline the calculation of the corresponding closure
for the heat flux in terms of the gradients of the independent scalar
state variables: number density and temperature. The results are compared
with the ones obtained using Marle's approximation in which a relativistic
correction factor is included for the relaxation time parameter. It
is shown how including such correction this BGK-like model yields
good results in the relativistic case and not only in the non-relativistic
limit as was previously found by other authors. 
\end{abstract}
\maketitle


%
\section{Introduction}
The interest in relativistic non-equilibrium thermodynamics has grown
recently, mostly due to recent experiments as well as new ideas in
theoretical modeling \cite{Kodama,Azwinndini1,Azwinndini2,Policastro}.
Most works dealing with applications of the theory do not refer to
the corresponding system of transport equations including dissipation\emph{
to first order in the gradients}. This is principally due to the fact
that for decades such system has been thought to predict unphysical
dynamics in the fluid \cite{Hiscock}. However, it was recently pointed
out that when kinetic theory is used to formulate the constitutive
equations that complete such set, instead of phenomenological arguments,
the system of transport equations is free of instabilities and shows
no causality issues in the linear regime \cite{PA2009}.

Relativistic kinetic theory to first order in the gradients has been
vastly explored by previous authors \cite{Israel,deGroot,kremer}.
In those works the purely relativistic term that appears in the closure
for the heat flux is included in a generalized thermal force with
one transport coefficient i.e. a relativistic thermal conductivity.
However, the system of transport equations constitutes a set of five
balance equations and needs to be eventually expressed in terms of
only five state variables. In the usual representation, these variables
are the number density $n$, the hydrodynamic four velocity $\mathcal{U}^{\nu}$,
which only has three independent components, and temperature $T$.
Because of this, constitutive equations for the dissipative fluxes
in terms of gradients of those quantities are required to close the
system. The aim of this work is to show the highlights of the calculation
of the heat flux from the complete Boltzmann equation using the Chapman-Enskog
approximation within the $n,\,\mathcal{U}^{\nu},\, T$ representation.
The details and further analysis of this calculation can be found
in Ref. \cite{aa2010}. In order to perform this task we consider
the density and temperature gradients as independent forces and write
two uncoupled integral equations. The coefficients involved in the
constitutive relation are calculated and plotted for a constant cross
section model. Here, we also compare the results with those found
in previous works using a BGK-like approximation \cite{PA2009,aip-proceedings2009}.
In order to make such comparison, we adjust the parameter included
in Marle's kernel by a relativistic factor arising from the structure
of the kinetic equation. We further show that when this factor is
taken into account, the deviation from the exact values are as significative
as in the non-relativistic case.

The rest of this work is organized as follows. In the second section some basic
elements of relativistic kinetic theory are briefly reviewed including
the Chapman-Enskog expansion for the solution of the Boltzmann equation.
In the third section an outline of the calculation of the heat flux using
two independent forces is shown. The results for the transport coefficients
are compared to the values obtained in the BGK approximation in section four and some final remarks and conclusions are included in the last section.
\section{Relativistic kinetic theory}
In the kinetic theory of gases, the evolution of the distribution
function of a a dilute gas is given by the Boltzmann equation. If
the temperature $T$ of such gas is high enough such that the relativistic
parameter $z=kT/mc^{2}$, where $m$ is the rest mass of the particles,
$c$ the speed of light and $k$ Boltzmann's constant, is not negligible,
the non-relativistic description is not appropriate. Instead, relativistic
kinetic theory (RKT) constitutes a more accurate framework. In RKT
the fact that the speed of the molecules can be close to the speed
of light is taken into account. The evolution of the distribution
function is then given by the special relativistic Boltzmann equation
\cite{lichnerowicz} 
\begin{equation}
v^{\alpha}f_{,\alpha}=J(f\, f'),\label{boltzmann}
\end{equation}
where $v^{\alpha}=\gamma\left(\omega^{\ell},c\right)$ is the molecular
four-velocity, $\gamma$ being Lorentz factor , and $J(f,f')$ is
the collision kernel \cite{kremer} 
\[J\left(f\, f'\right)=\int\int\left[f\,'f_{1}\,'-f\, f_{1}\right]\mathcal{F}\sigma\left(\Omega\right)d\Omega dv_{1}^{*}\]
In the previous equation $\mathcal{F}$ the invariant flux and $\sigma\left(\Omega\right)d\Omega$
is the differential cross section \cite{kremer}. The system
here considered is a dilute, single component, neutral, non-degenerate,
inert gas. The local variables, number density $n$, hydrodynamic
four-velocity $u^{\nu}$ and internal energy density $e$, are defined
as \begin{equation}
n=\int f\gamma dv^{*},\label{ene}
\end{equation}
\begin{equation}
nu^{\nu}=\int fv^{\nu}dv^{*},\label{vel-hidro}
\end{equation}
\begin{equation}
ne=mc^{2}\int f\gamma^{2}dv^{*}.\label{energia}
\end{equation}
respectively. The local equilibrium distribution function, which is
the solution of the homogeneous Boltzmann equation, in the relativistic
case is given by Ref. \cite{Juttner} 
\begin{equation}
f^{(0)}=\frac{n}{4\pi c^{3}}\frac{1}{zK_{2}\left(\frac{1}{z}\right)}\exp^{\frac{u^{\nu}v_{\nu}}{zc^{2}}}\label{eq:jutner}
\end{equation}
where $K_{n}(x)$ is the modified Bessel function of the second kind.
As shown in Ref. \cite{AAL2010}, the heat flux can be calculated
as the average of the chaotic kinetic energy, that is\begin{equation}
q^{\mu}=mc^{2}h_{\nu}^{\mu}\int\gamma v^{\nu}fd^{*}v\label{eq:qq}\end{equation}
The chaotic velocity corresponds to the velocity measured in a local
comoving frame. Thus, one can replace the invariant $u^{\nu}v_{\nu}$
in Eq. (\ref{eq:jutner}) by its value in the local comoving frame,
that is $u^{\nu}v_{\nu}=-c^{2}\gamma$. In this context the total
energy and momentum fluxes for an arbitrary observer are given by
the Lorentz transformed dissipative fluxes as calculated in the local
comoving frame. This argument is clearly discussed in Refs. \cite{aa2010,AAL2010}.

In order to calculate the heat flux, given in Eq. (\ref{eq:qq}),
one does not need an explicit solution to the Boltzmann equation.
Using Hilbert's method one can obtain a good approximation to the
integral in Eq. (\ref{eq:qq}). Following the prescription of such
method, the velocity distribution function is written as $f=f^{(0)}+f^{(1)}$
where $f^{\left(1\right)}$ is considered to be a first order correction
in the gradients of the state variables. Also, in order to associate
the local variables to the equilibrium state, the following subsidiary
conditions need to be imposed 
\begin{eqnarray}
\int f^{(1)}\left(\begin{array}{c}
\gamma^{2}\\
v^{\nu}\end{array}\right)dv^{*}=0.\label{eq:sub}
\end{eqnarray}
Under these hypotheses, the linearized relativistic Boltzmann equation,
obtained by substituting $f$ in Eq. (\ref{boltzmann}) and considering
only linear deviations from local equilibrium, reads 
\begin{equation}
v^{\alpha}f_{,\alpha}^{(0)}=f^{(0)}\mathbin{C}(\phi)\label{eq:lbe}
\end{equation}
where 
\[\mathbin{C}\left(\phi\right)=\int\int\left\{ \phi'_{1}+\phi'-\phi_{1}-\phi\right\} f_{1}^{\left(0\right)}\mathcal{F}\sigma\left(\Omega\right)d\Omega dv_{1}^{*},\]
is the linearized collision kernel. Substituting $f^{(0)}$ on the
left hand side and using the identity $v^{\alpha}A_{,\alpha}=h^{\mu\nu}v_{\nu}A_{,\mu}-\frac{\mathcal{U}^{\mu}\mathcal{U}^{\nu}}{c^{2}}v_{\nu}A_{,\mu}$ Eq.
(\ref{eq:lbe}) yields

\begin{eqnarray}
v^{\alpha}f_{,\alpha}^{(0)} & =f^{(0)}h^{\mu\nu}v_{\nu}\left(\frac{n_{,\mu}}{n}-\frac{T_{,\mu}}{T}\left[1+\frac{u^{\beta}v_{\beta}}{zc^{2}}+\frac{1}{2z}\left(\frac{K_{1}(\frac{1}{z})+K_{3}(\frac{1}{z})}{K_{2}(\frac{1}{z})}\right)\right]+\frac{v_{\beta}u_{;\mu}^{\beta}}{zc^{2}}\right)\nonumber \\
 & +f^{(0)}\gamma c\left(\frac{\dot{n}}{n}-\frac{\dot{T}}{T}\left[1+\frac{u^{\beta}v_{\beta}}{zc^{2}}+\frac{1}{2z}\left(\frac{K_{1}(\frac{1}{z})+K_{3}(\frac{1}{z})}{K_{2}(\frac{1}{z})}\right)\right]+\frac{v^{\beta}\dot{u}_{\beta}}{zc^{2}}\right).\label{left}
\end{eqnarray}
Notice that the first line in Eq. (\ref{left}) only includes the
spatial components, since in the comoving frame $h^{\mu4}=0$ while
the time derivatives are isolated in the second line which we have
simplified using that
\[\frac{u^{\mu}u^{\nu}}{c^{2}}v_{\nu}=-u^{\mu}\gamma\delta_{\mu}^{4}\]
This separation between temporal and spatial parts is particularly
useful since for the Chapman-Enskog solution to first order in the
gradients to exist, the previous order equations have to be satisfied.
Because of this, the relativistic Euler equations namely, 
\[\dot{n}=-nu_{;\alpha}^{\alpha},\]
\begin{equation}
\dot{u}=-\left(\frac{ne}{c^{2}}+\frac{p}{c^{2}}\right)^{-1}p_{,\mu}h_{\beta}^{\mu},\label{euler}
\end{equation}
\[\dot{T}=-\frac{T\beta}{nC_{n}k}u_{;\alpha}^{\alpha},\]
are substituted in the expression in Eq. (\ref{left}) such that only
spatial gradients appear in it. Also, since the momentum balance equation
is in terms of the gradient of the hydrostatic pressure, we introduce
in it the relation 
\begin{equation}
p_{,\mu}=\left(\frac{\partial p}{\partial n}\right)n_{,\mu}+\left(\frac{\partial p}{\partial T}\right)T_{,\mu}.
\end{equation}
in order to be consistent with our representation. After Euler Eqs. 
(\ref{euler}), are introduced in Eq. (\ref{left}) and using the equation
state for an ideal gas $p=nkT$, we can write 
\begin{equation}
v^{\beta}h_{\beta}^{\alpha}\left\{ \left(1-\gamma\frac{K_{2}(\frac{1}{z})}{K_{3}(\frac{1}{z})}\right)\frac{n_{,\alpha}}{n}+\frac{T_{,\alpha}}{T}\left(1+\frac{\gamma}{z}-\gamma\frac{K_{2}(\frac{1}{z})}{K_{3}(\frac{1}{z})}-\frac{K_{3}(\frac{1}{z})}{zK_{2}(\frac{1}{z})}\right)\right\} =\mathbin{C}(\phi).\label{inhomogenea1}
\end{equation}
Since in this work we are only interested in the heat flux, we have
ignored the term proportional to $u_{;\alpha}^{\beta}$ in Eq. (\ref{inhomogenea1})
due to Curie's principle.

Following Hilbert's method, the general solution for the integral
equation (\ref{inhomogenea1}) is given by the sum of a particular
solution and a linear combination of the collision invariants as solutions
to the homogeneous equation. That is 
\begin{equation}
\phi=\mathcal{A}(\gamma)v^{\alpha}h_{\beta}^{\alpha}\frac{T_{,\alpha}}{T}+\mathcal{B}(\gamma)v^{\alpha}h_{\beta}^{\alpha}\frac{n_{,\alpha}}{n}+\alpha+\tilde{\alpha}_{\nu}v^{\nu}\label{soluciongeneral}\end{equation}
where $\mathcal{A}(\gamma)$ and $\mathcal{B}(\gamma)$ are functions
of $\gamma$ and the local variables $n,\vec{v},T$ while $\alpha$
and $\tilde{\alpha}_{\nu}$ are constants. By imposing the subsidiary
conditions, required for the uniqueness of the solution, Eq. (\ref{eq:sub}),
$\phi$ can be rewritten as 
\begin{equation}
\phi=\mathcal{A}(\gamma)v^{\beta}h_{\beta}^{\alpha}\frac{T_{,\alpha}}{T}+\mathcal{B}(\gamma)v^{\beta}h_{\beta}^{\alpha}\frac{n_{,\alpha}}{n}.\label{sol-gen-sub}
\end{equation}
The coefficients $\mathcal{A}(\gamma)$ and $\mathcal{B}(\gamma)$
are further expressed in terms of orthogonal polynomials in the form
\begin{eqnarray*}
\mathcal{A}(\gamma)=\sum_{n=0}^{\infty}a_{n}\mathcal{L}_{n}(\gamma),\\
\mathcal{B}(\gamma)=\sum_{n=0}^{\infty}b_{n}\mathcal{L}_{n}(\gamma),\end{eqnarray*}
with $\mathcal{L}_{n}(\gamma)=\sum_{k=0}^{n}\alpha_{kn}\gamma^{k}$.
The elements of this set satisfy the orthogonality condition \[
\int\mathcal{L}_{n}(\gamma)\mathcal{L}_{m}(\gamma)\exp^{-\frac{\gamma}{z}}(\gamma^{2}-1)^{3/2}d\gamma=\delta_{nm},\]
More details on these polynomials can be consulted in Refs. \cite{kremer,aa2010}.
Direct substitution of coefficients $\mathcal{A}(\gamma)$ and $\mathcal{B}(\gamma)$
in the solution (\ref{sol-gen-sub}) leads us to the expression \begin{equation}
\phi=h_{\beta}^{\alpha}\sum_{n=0}^{\infty}a_{n}\mathcal{L}_{n}(\gamma)v^{\beta}{\frac{T_{,\alpha}}{T}}+h_{\beta}^{\alpha}\sum_{n=0}^{\infty}b_{n}\mathcal{L}_{n}(\gamma)v^{\beta}{\frac{n_{,\alpha}}{n}},\label{sol-fin}\end{equation}
It is worthwhile pointing out that the first-order approximation for
the distribution function is obtained here in terms of the density
and temperature gradients as was already discussed in \cite{PA2009,aa2010}.
This is because of the chosen representation in which $n$ and $T$ are
considered independent state variables. Thus, when the heat flux is
calculated from the distribution function, the term proportional to
the temperature gradient will lead us to a Fourier-like law, while
the term proportional to the density gradient is a purely relativistic
contribution. The latter has no non-relativistic counterpart since
the corresponding term in Eq. (\ref{inhomogenea1}) vanishes in such
limit.

Substituting the proposed solution $\phi$ in the linearized Boltzmann
equation (\ref{inhomogenea1}) leads to an integral equation 
\[v^{\beta}h_{\beta}^{\alpha}\left\{ \left(1-\gamma\frac{K_{2}(\frac{1}{z})}{K_{3}(\frac{1}{z})}\right)\frac{n_{,\alpha}}{n}+\frac{T_{,\alpha}}{T}\left(1+\frac{\gamma}{z}-\gamma\frac{K_{2}(\frac{1}{z})}{K_{3}(\frac{1}{z})}-\frac{K_{3}(\frac{1}{z})}{zK_{2(\frac{1}{z})}}\right)\right\} =\mathbin{C}(\phi).\]
which can be separated into two independent equations as follows \[
v^{\beta}h_{\beta}^{\alpha}\left\{ 1+\frac{\gamma}{z}-\gamma\frac{K_{2}(\frac{1}{z})}{K_{3}(\frac{1}{z})}-\frac{K_{3}(\frac{1}{z})}{zK_{2}(\frac{1}{z})}\right\} =h_{\beta}^{\alpha}\sum_{n=1}^{\infty}a_{n}\mathbin{C}(\mathcal{L}_{n}(\gamma)v^{\beta})\]

\[v^{\beta}h_{\beta}^{\alpha}\left\{ 1-\gamma\frac{K_{2}(\frac{1}{z})}{K_{3}(\frac{1}{z})}\right\} =h_{\beta}^{\alpha}\sum_{n=1}^{\infty}b_{n}\mathbin{C}(\mathcal{L}_{n}(\gamma)v^{\beta}).\]
As mentioned above, the explicit complete solution for $f^{\left(1\right)}$ is not required
in order to calculate the heat flux. As will be shown in the next
section only one term in the infinite sums appearing in the previous
equations is enough in order to get a good approximation for $q^{\mu}$.

\section{Heat flux}
The heat flux is physically the average
of a chaotic energy flux. Because of this, it is calculated in the local comoving frame, as is clearly stated in Eq. (\ref{eq:qq}). Substituting $f^{(1)}=f^{(0)}\phi$
with $\phi$ given by Eq. (\ref{sol-fin}) in Eq. (\ref{eq:qq}) yields
\begin{equation}
q^{\mu}=\frac{nm}{4\pi czK_{2}(\frac{1}{z})}\left[I_{a}^{\mu}+I_{b}^{\mu}\right],\end{equation}
where the integrals in the bracket are given by\begin{equation}
I_{a}^{\nu}=\int\left[\mathcal{A}(\gamma)v_{k}\frac{T^{,k}}{T}\right]\gamma v^{\nu}\exp^{-\frac{\gamma}{z}}\sqrt{\gamma^{2}-1}d\gamma,\label{eq:i1}\end{equation}
\begin{equation}
I_{b}^{\nu}=\int\left[\mathcal{B}(\gamma)v_{k}\frac{n^{,k}}{n}\right]\gamma v^{\nu}\exp^{-\frac{\gamma}{z}}\sqrt{\gamma^{2}-1}d\gamma.\label{eq:i2}\end{equation}
 In Ref. \cite{aa2010} it is shown how, of all coefficients $a_{n}$
and $b_{n}$ in Eqs. (\ref{eq:i1}) and (\ref{eq:i2}), only one of
each is required in order to obtain a good first approximation to
$I_{a}^{\nu}$ and $I_{b}^{\nu}$. After a cumbersome calculation,
which can be found to some detail in Ref. \cite{aa2010}, and assuming
constant collisions cross-section (see \cite{kremer}) one obtains
\begin{equation}
q^{\mu}=-h^{\mu\alpha}\left[L_{T}\frac{T_{,\alpha}}{T}+L_{n}\frac{n_{,\alpha}}{n}\right],\label{heat-flux}\end{equation}
where the transport coefficients $L_{T}$ and $L_{n}$ can be written
as \begin{equation}
L_{T}=\frac{3}{64}\frac{ckT}{\pi\sigma}f_{1}(z)\qquad L_{n}=\frac{3}{64}\frac{ckT}{\pi\sigma}f_{2}(z)\label{LT}\end{equation}
where \begin{equation}
f_{1}(z)=\left(1-z\frac{K_{2}(\frac{1}{z})}{K_{3}(\frac{1}{z})}\right)\frac{\left(K_{2}\left(\frac{1}{z}\right)\left[\frac{1}{z}+5\mathcal{G}\left(\frac{1}{z}\right)-\frac{1}{z}\mathcal{G}\left(\frac{1}{z}\right)^{2}\right]\right)^{2}}{z^{4}\left(\frac{5}{z}K_{3}\left(\frac{2}{z}\right)+\left(\frac{1}{z^{2}}+2\right)K_{2}\left(\frac{2}{z}\right)\right)}\label{l1}\end{equation}
 \begin{equation}
f_{2}(z)=-\frac{K_{2}\left(\frac{1}{z}\right)\left(K_{2}\left(\frac{1}{z}\right)\left[\frac{1}{z}+5\mathcal{G}\left(\frac{1}{z}\right)-\frac{1}{z}\mathcal{G}\left(\frac{1}{z}\right)^{2}\right]\right)^{2}}{z^{3}K_{3}\left(\frac{1}{z}\right)\left(\frac{5}{z}K_{3}\left(\frac{2}{z}\right)+\left(\frac{1}{z^{2}}+2\right)K_{2}\left(\frac{2}{z}\right)\right)}.\label{l2}\end{equation}
 and $\mathcal{G}(1/z)=K_{3}(1/z)/K_{2}(1/z)$. In order to compare
these results with the ones obtained in the relaxation time approximation
we will explore Marle's model in the next section and find appropriate
expressions for the functions $f_{1}(z)$ and $f_{2}(z)$ in such
context. 

\section{Comparison with the BGK approximation}
The BGK approximation consists in replacing the right hand side of
the Boltzmann equation with a simplified kernel in which all the details
of the collisions are substituted by a parameter. The kinetic equation
within relaxation models reads 
\begin{equation}
v^{\alpha}f_{,\alpha}=-\mathcal{K}(f-f^{(0)}).\label{marle}
\end{equation}
where the parameter $\mathcal{K}$ is assumed to be independent of
$v^{\mu}$. The transport coefficients obtained using this method
will have a strong dependence on this parameter. Thus, in order to
compare the results obtained in the previous section using the complete
kernel with the ones obtained with this model we need an approximation
for the parameter $\mathcal{K}$. 

To understand the role of this constant we follow the standard reasoning.
We assume that the one particle distribution function does not depend
on the spatial coordinates such that we can write \[
\gamma\frac{df}{dt}=-\mathcal{K}(f-f^{(0)})\]
from where one obtains 
\begin{equation}
f(t)=\exp\left(-\frac{\mathcal{K}t}{\gamma}\right)\left[f(0)+\frac{\mathcal{K}}{\gamma}\int_{0}^{t}\exp\left(\frac{\mathcal{K}x}{\gamma}\right)f^{(0)}(x)dx\right]\label{eq:fbgk}
\end{equation}
Equation (\ref{eq:fbgk}) motivates the association of the quantity
$\gamma/\mathcal{K}$ with a characteristic relaxation time $\tau$.
In the non-relativistic case, the analog to Eq. (\ref{eq:fbgk}) leads
to assuming $\tau\sim\mathcal{K}$. In Ref. \cite{kremer} the
$\gamma$ factor is not included in $\mathcal{K}$ which at the end
leads to results for the thermal conductivity that deviate substantially
from the complete solution value in the relativistic and ultra-relativistic
scenarios. Only when $\gamma\to1$, in the non-relativistic limit,
the results are consistent.

Here we consider $\tau\sim\gamma/\mathcal{K}$ and propose a reasonable
estimate for the parameter $\mathcal{K}$ from known characteristic
times related to the collisions in the system. In non-relativistic
kinetic theory, the time between collisions $\tau_{c}$ for hard-sphere
particles of cross section $\sigma$ is given by 
\[\tau_{c}=\frac{1}{4n\pi\sigma\langle g\rangle}\]
where $\langle g\rangle$ denotes the mean value of the relative velocity
of two particles. Usually, the relaxation
time $\tau$ appearing in BGK model, is considered proportional to
$\tau_{c}$ and since $\langle g\rangle$ is of the order of the mean
velocity $\langle v\rangle$ and/or the adiabatic sound speed $v_{s}=\sqrt{5kT/3m}$,
we propose for this case \begin{equation}
\frac{\gamma}{\mathcal{K}}=\frac{1}{4n\pi\sigma V(z)}\label{k-marle}\end{equation}
 where $V(z)$ is the characteristic speed proportional to $\langle v\rangle$
or $v_{s}$. Finally to approximate $\mathcal{K}$ we propose \begin{equation}
\frac{1}{\mathcal{K}}\sim\frac{1}{4n\pi\sigma V(z)\langle\gamma\rangle},\label{k-marle-propuesta}\end{equation}
 in analogy with the non-relativistic case.

Using Marle's model, Sandoval \textit{et al.} \cite{PA2009} obtained
expressions for the transport coefficients which can be written in
same fashion as Eq. (\ref{LT}) where the functions $f_{1}(z)$ and
$f_{2}(z)$ are replaced by \begin{equation}
f_{1M}(z)=\frac{16c}{3}\left(1-z\frac{K_{2}}{K_{3}}\right)\left\{ \frac{1}{z}+5\mathcal{G}\left(\frac{1}{z}\right)-\frac{1}{z}\mathcal{G}\left(\frac{1}{z}\right)^{2}\right\} \frac{4n\pi\sigma}{\mathcal{K}}\label{eq:f1m}\end{equation}
 \begin{equation}
f_{2M}(z)=-\frac{16c}{3}\frac{zK_{2}}{K_{3}}\left\{ \frac{1}{z}+5\mathcal{G}\left(\frac{1}{z}\right)-\frac{1}{z}\mathcal{G}\left(\frac{1}{z}\right)^{2}\right\} \frac{4n\pi\sigma}{\mathcal{K}}.\label{eq:f2m}\end{equation}
 where we have included the $M$'s subscript to indicate Marle's model
solution. %
\begin{figure}
\noindent \begin{centering}
\includegraphics[scale=0.55]{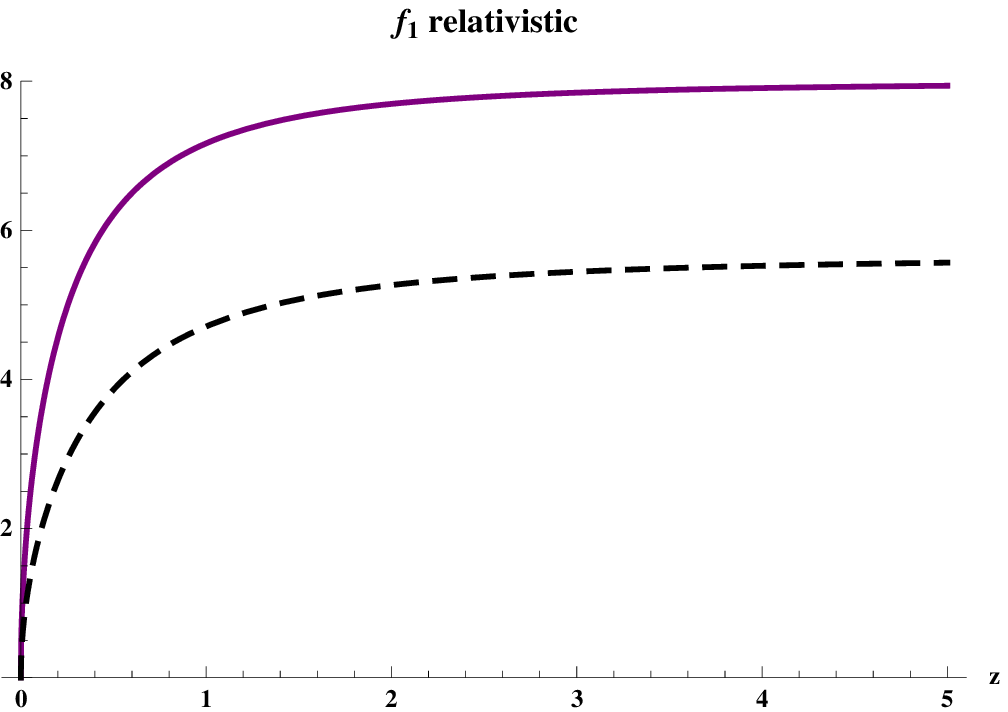}
\hspace{0.5cm}
\includegraphics[scale=0.7]{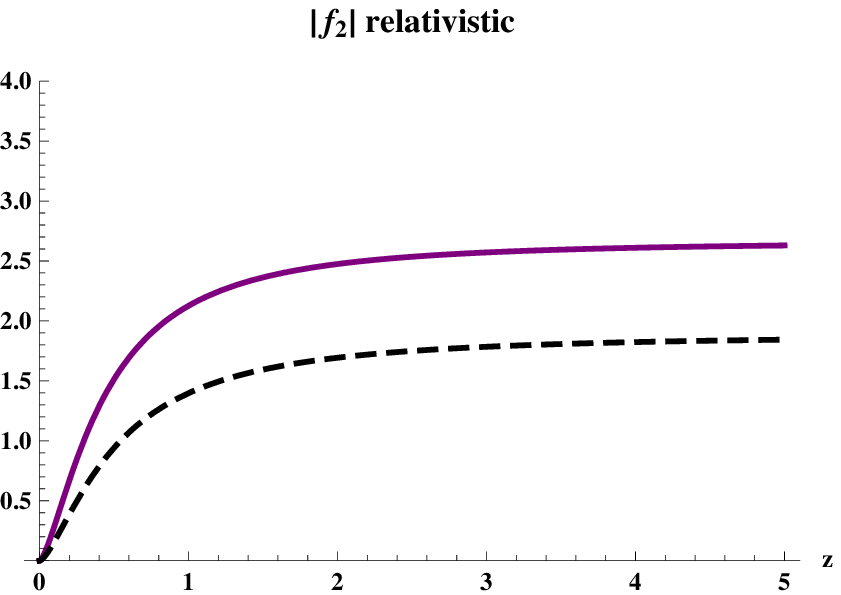} 
\par\end{centering}
\caption{Relativistic coefficients $f_{1}$ and $f_{2}$. The continuous line
for the complete kernel and the dashed line for Marle's model including the $\gamma$ factor in the relaxation parameter.}\label{fig1}
\end{figure}

\begin{figure}
\noindent \begin{centering}
\includegraphics[scale=0.7]{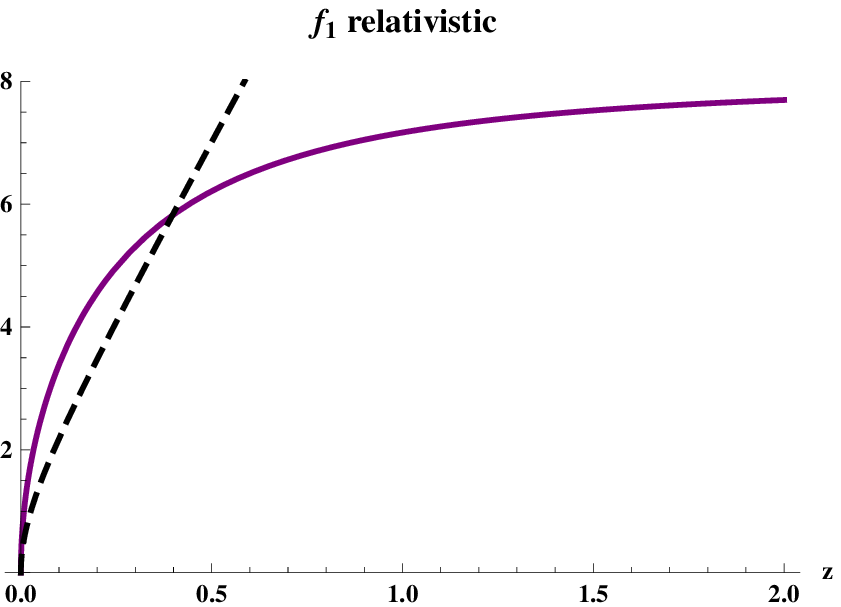}
\hspace{0.5cm}
\includegraphics[scale=0.7]{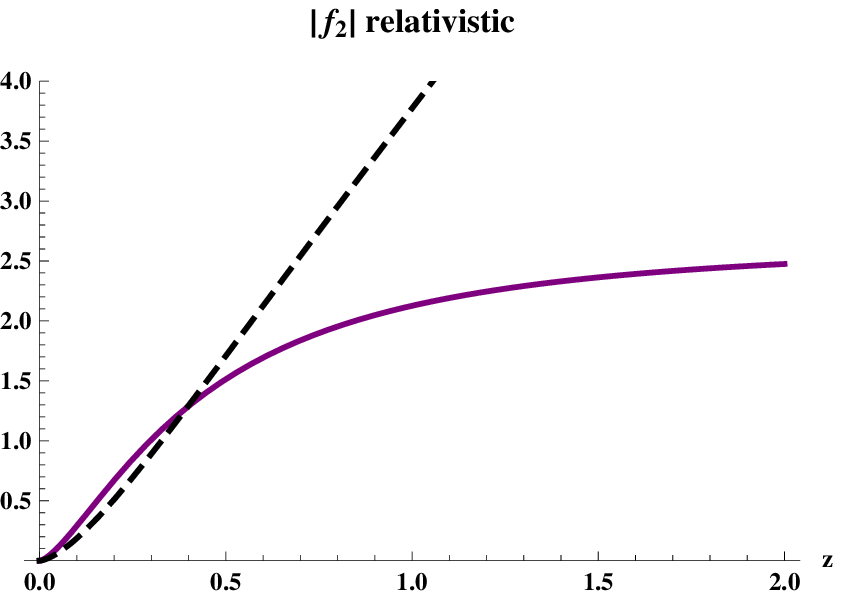} 
\par\end{centering}
\caption{Relativistic coefficients $f_{1}$ and $f_{2}$. The continuous line for the complete kernel and the dashed line for Marle's model without taking into account the $\gamma$ factor in the relaxation parameter.}\label{fig2}
\end{figure}

\begin{figure}
\includegraphics[scale=0.75]{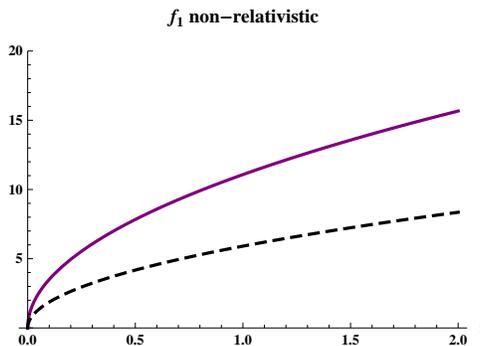} 
\caption{Non-relativistic case with the complete kernel (continuous line) and
the BGK model (dashed line). Only the thermal conductivity factor
$f_{1}$ is shown since $f_{2}$ corresponds to a purely relativistic
correction.}\label{fig3}
\end{figure}

The results of substituting expression (\ref{k-marle}) into (\ref{eq:f1m})
and (\ref{eq:f2m}) are shown in Fig. \ref{fig1}, where we have considered
$V(s)=\sqrt{2}\langle v\rangle$. As can be clearly seen from the
plot, the deviation of the approximated solution from the exact one
is not as significant as without the $\gamma$ factor, which is shown
in Fig. \ref{fig2}. Moreover, the difference between exact and approximated
solutions when the $\gamma$ factor is taken into account is not grater than
to the one obtained in the non-relativistic case which is shown in
Fig. \ref{fig3}, for the sake of comparison.

With the information in hand, one can also infer the most reasonable
expression for the characteristic velocity $V(z)$ with the help of
both models. To accomplish this task, we write \[
\frac{1}{\mathcal{K}}\frac{L_{T}}{L_{MT}}=\frac{3}{64\pi\sigma nc}\frac{K_{2}\left(\frac{1}{z}\right)^{2}\left[\frac{1}{z}+5\mathcal{G}\left(\frac{1}{z}\right)-\frac{1}{z}\mathcal{G}\left(\frac{1}{z}\right)^{2}\right]}{z^{4}\left(\frac{5}{z}K_{3}\left(\frac{2}{z}\right)+\left(\frac{1}{z^{2}}+2\right)K_{2}\left(\frac{2}{z}\right)\right)}\]
 and \[
\frac{1}{\mathcal{K}}\frac{L_{n}}{L_{Mn}}=\frac{3}{64\pi c\sigma n}\frac{K_{2}\left(\frac{1}{z}\right)^{2}\left[\frac{1}{z}+5\mathcal{G}\left(\frac{1}{z}\right)-\frac{1}{z}\mathcal{G}\left(\frac{1}{z}\right)^{2}\right]}{z^{4}\left(\frac{5}{z}K_{3}\left(\frac{2}{z}\right)+\left(\frac{1}{z^{2}}+2\right)K_{2}\left(\frac{2}{z}\right)\right)}\]
and consider the best approximation to $V(z)$ to be the one that
makes both coefficients $L_{T}$'s or $L_{n}$'s match perfectly. The
results are shown in Fig. 4 together with the characteristic velocity
given by Eq. (\ref{k-marle-propuesta}) and $v_{s}$. Note that $V_{c}\left(z\right)$
is the value one should give to the characteristic relative speed
for the relaxation time model to match exactly the results obtained
with the complete kernel. %
\begin{figure}
\includegraphics[scale=0.65]{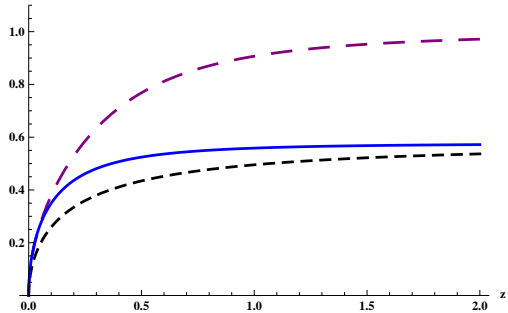} 
\caption{This plot shows a comparison of the different characteristic velocities
$V_{c}(z)$ (long dashed line), $\langle v\rangle(z)$ (dashed line) and $v_{s}(z)$ (continuous line) .}
\end{figure}

\section{Final remarks}
The relativistic hydrodynamic equations at the Navier-Stokes level
need closure relations in order to constitute a complete set. In particular,
the constitutive equation for the heat flux can be written in terms
of the gradients of the temperature and the density when such quantities
are considered independent state variables. The corresponding coupling
coefficients have been calculated before assuming Marle's model \cite{PA2009,aip-proceedings2009}.
Here we outlined the procedure for the exact calculation, within Chapman-Enskog's
approximation, and compared the results in the case of constant cross-section
collisions.

We re-analyzed Marle's proposal as a relaxation model and found that
the quantity related to a relaxation time is actually best approximated
by $\langle\gamma\rangle/\mathcal{K}$. With this correction Marle's
model proves to be a good approximation not only in very mildly relativistic
systems. Additionally we compare the different values that can be
assigned to the characteristic microscopic velocity when approximating
the value for $\mathcal{K}$ with the one inferred from the exact
solution.

We finally wish to point out the similarity in structure of the second
term in Eq. (\ref{heat-flux}) with the Dufour effect. This is only
in the sense that such relation implies the existence of a heat flux
even in isothermic conditions due to a gradient in density. However,
the coupling between heat and the density gradient in relativistic
gases cannot be properly referred to as a cross effect since there
is no reciprocal (diffusive) flux. The Dufour effect is the heat flux
due to a diffusive force that appears in more than one component systems
and has a conjugate effect called Soret effect. The role of the term
here studied in the case of a binary mixture is still not clear and
will be addressed in the future.
\vspace{1cm}
 
\noindent \textbf{Acknowledgements:}
This work was partially supported by PROMEP grant UAM-PTC-142. 
\bibliographystyle{aipproc}

\end{document}